\documentclass{article}

\usepackage[margin=1in]{geometry}
\usepackage{graphicx}
\usepackage{amsmath}
\usepackage{amssymb} 
\usepackage{booktabs}
\usepackage{enumitem}
\usepackage{natbib} 
\usepackage{hyperref}
\hypersetup{
    colorlinks=true,
    linkcolor=blue,
    filecolor=magenta,
    urlcolor=cyan,
    citecolor=blue
}
\usepackage[T1]{fontenc} 
\usepackage{lmodern}     
\usepackage[english]{babel} 
\usepackage{microtype} 

\title{Interpretable Machine Learning for Macro Alpha: A News Sentiment Case Study}

\author{
    Yuke Zhang\thanks{This research was conducted in the author's personal capacity. The views expressed herein are solely those of the author and do not necessarily represent the views of any institution. The code for this research is available at: \url{https://github.com/yukepenn/macro-news-sentiment-trading}.} \\ 
    \texttt{\href{mailto:yukez@seas.upenn.edu}{yukez@seas.upenn.edu}} 
}
\date{\today}

\begin{document}

\maketitle

\begin{abstract}
This study introduces an interpretable machine learning (ML) framework to extract macroeconomic alpha from global news sentiment. We process the Global Database of Events, Language, and Tone (GDELT) Project's worldwide news feed using FinBERT—a Bidirectional Encoder Representations from Transformers (BERT) based model pretrained on finance-specific language—to construct daily sentiment indices incorporating mean tone, dispersion, and event impact. These indices drive an XGBoost classifier, benchmarked against logistic regression, to predict next-day returns for EUR/USD, USD/JPY, and 10-year U.S.\ Treasury futures (ZN). Rigorous out-of-sample (OOS) backtesting (5-fold expanding-window cross-validation, OOS period: c. 2017–April 2025) demonstrates exceptional, cost-adjusted performance for the XGBoost strategy: Sharpe ratios achieve 5.87 (EUR/USD), 4.65 (USD/JPY), and 4.65 (Treasuries), with respective compound annual growth rates (CAGRs) exceeding 50\% in Foreign Exchange (FX) and 22\% in bonds. Shapley Additive Explanations (SHAP) affirm that sentiment dispersion and article impact are key predictive features. Our findings establish that integrating domain-specific Natural Language Processing (NLP) with interpretable ML offers a potent and explainable source of macro alpha.
\end{abstract}


\section{Introduction}

In an era of rapid information diffusion, financial markets react instantaneously to news and shifts in investor sentiment. This phenomenon is particularly evident in macro assets—such as foreign exchange (FX) and fixed income—where unstructured text from press releases, central bank announcements, and geopolitical reports can influence significant capital flows. While these rich data sources promise novel trading insights, their sheer volume and linguistic complexity pose substantial challenges. Early sentiment-based strategies often relied on coarse lexicons, frequently misclassifying domain-specific terms \citep{loughran2011liability}. The advent of finance-trained language models has transformed this landscape: FinBERT \citep{araci2019finbert}, a Bidirectional Encoder Representations from Transformers (BERT) variant, achieves state-of-the-art accuracy in extracting nuanced sentiment from financial texts.

Concurrently, quantitative trading has embraced machine learning (ML) models, particularly gradient-boosted trees and deep networks, which excel at capturing nonlinear interactions among features. Yet, this predictive power often comes at the cost of transparency, a critical shortcoming in regulated environments where model explainability underpins risk management and compliance. Shapley Additive Explanations (SHAP) \citep{lundberg2017shap} offer a principled solution by assigning each input feature a consistent and theoretically sound contribution to individual predictions, thereby demystifying "black-box" algorithms.

This paper integrates these advances—domain-specific Natural Language Processing (NLP), cross-asset macro data, and interpretable ML—to investigate whether global news sentiment, processed with modern tools and explained transparently, can generate systematic alpha in macro trading. We target three representative markets: the EUR/USD and USD/JPY spot FX rates, and the 10-year U.S.\ Treasury futures (ZN). Our pipeline ingests daily news from the Global Database of Events, Language, and Tone (GDELT) Project, applies FinBERT to score each article’s sentiment, and aggregates these scores into a suite of features (including mean tone, dispersion, volume, and Goldstein scores, along with their lags and rolling statistics). We then train two classifiers—a logistic regression (as a transparent baseline) and an XGBoost model \citep{chen2016xgboost} (to exploit nonlinearities)—to predict next-day return direction.

Our contributions are fourfold:
\begin{enumerate}[noitemsep]
  \item \textbf{Cross-Asset News Sentiment Trading Strategy:} We demonstrate that a unified news-driven strategy can produce significant risk-adjusted returns across both FX and bond markets, extending literature that often focuses on equities \citep{filippou2020fxnews, ravenpack2023fx}.
  \item \textbf{Advanced NLP Integration:} By leveraging FinBERT, we aim to improve sentiment signal quality over traditional lexicons, capturing context-dependent nuances in economic news.
  \item \textbf{Model Interpretability:} Employing SHAP, we dissect the XGBoost model’s decisions, confirming that features such as sentiment dispersion and article impact drive predictions in line with economic intuition.
  \item \textbf{Robust Out-of-Sample (OOS) Evidence:} Our backtesting framework uses data from January 2015 through April 2025 with an expanding window methodology for OOS evaluation. Including realistic transaction costs, the strategy yields impressive OOS Sharpe ratios: 5.87 (EUR/USD), 4.65 (USD/JPY), and 4.65 (ZN 10-year Treasury futures), with results persisting across subperiods and cost assumptions.
\end{enumerate}

Beyond demonstrating strong empirical performance, our work underscores the feasibility of developing transparent and reproducible macro trading strategies built on publicly available data and open-source models.

The remainder of this paper is organized as follows. Section~\ref{sec:lit_review} provides a review of related literature and highlights our originality. Section~\ref{sec:data_method} details our data sources, sentiment extraction methodology, and feature engineering process. Section~\ref{sec:results_analysis} presents the predictive modeling and trading strategy results, alongside SHAP-based interpretability analyses. Section~\ref{sec:discussion} discusses cross-asset differences, robustness checks, and broader implications. Finally, Section~\ref{sec:conclusion} concludes and outlines avenues for future research.

\section{Literature Review and Originality}
\label{sec:lit_review}

To contextualize our contributions, we review key studies in news sentiment trading and NLP applications in macro finance, thereby highlighting the novelty of our approach.

\begin{itemize}[leftmargin=*]
  \item \textbf{FX News Sentiment Strategies:} \citet{filippou2020fxnews} examined news articles related to FX, constructing a sentiment-based currency strategy. They found a contrarian effect, where currencies with low media sentiment outperformed those with high sentiment. Our project also targets FX but employs an advanced sentiment model (FinBERT) and incorporates time-series predictive modeling with interpretable ML, extending the analysis to bond markets.

  \item \textbf{Macroeconomic News \& Interest Rates:} \citet{audrino2024macronews} demonstrated that sentiment extracted from macroeconomic news (specifically articles about interest rates, inflation, and labor markets) has significant explanatory power for short-term Treasury yield movements. Our project aligns with their findings by targeting rate markets but focuses on a daily global news flow via GDELT and a trading strategy perspective, complemented by SHAP for interpretability.

  \item \textbf{RavenPack FX Sentiment:} A white paper by \citet{ravenpack2023fx} showed that macro news sentiment indicators can predict G7 currency movements. While this underscores the value of news sentiment for FX, our project is distinct in its use of an open data source (GDELT) and transparent methods (FinBERT + XGBoost), adding a rigorous interpretability layer and consideration of bond markets.

  \item \textbf{Large Language Models (LLMs) and Financial News:} \citet{kirtac2024llm} investigated LLMs for sentiment trading on U.S.\ stock news, finding that GPT-3 outperformed FinBERT for equities. Their work highlights the evolving NLP landscape. Our study uses FinBERT, which is well-suited for our objectives and publicly available, but acknowledges that future work could incorporate newer or larger LLMs. Our focus remains distinct: macro assets (FX, rates) and global news.

  \item \textbf{GDELT in Macro Forecasting:} \citet{tilly2021gdelt} used GDELT news data to improve macroeconomic forecasts by training a Bidirectional Long Short-Term Memory (BiLSTM) network to classify emotions in global news. We build on a similar ethos (global data, richer sentiment measures) but pivot to trading signals and a different toolset (FinBERT for sentiment, XGBoost for prediction), focusing on direct market impact.
\end{itemize}

\subsection*{How Our Work Differs}
Our work distinguishes itself through the following combination of elements:
\begin{itemize}[noitemsep]
  \item \textbf{Advanced Domain-Specific NLP:} We employ FinBERT \citep{araci2019finbert}, a finance-domain pretrained transformer, for more accurate, context-aware sentiment signals compared to simpler lexicons or general-purpose models.
  \item \textbf{Global Coverage and Targeted Filtering:} Our pipeline utilizes GDELT’s comprehensive global news feed, specifically filtered for macroeconomic relevance, enabling broader event capture.
  \item \textbf{Integrated Cross-Asset Framework:} We apply a unified methodological approach to multiple asset classes (FX and bond futures), demonstrating the generality of news sentiment signals across diverse macro markets.
  \item \textbf{Prioritized Interpretability:} We employ SHAP \citep{lundberg2017shap} to explain model predictions, providing crucial insights into the decision-making process—an aspect often underemphasized in performance-driven studies.
  \item \textbf{Emphasis on Reproducibility:} Our approach is built on publicly available data (GDELT) and open-source models (FinBERT, XGBoost), rendering the research transparent and facilitating replication by other researchers.
\end{itemize}
To our knowledge, no prior publication has combined FinBERT with GDELT to produce interpretable trading signals for both FX and bond markets using a robust ML framework that demonstrates this level of out-of-sample performance.

\section{Data and Methodology}
\label{sec:data_method}
Our methodology encompasses news collection and filtering, sentiment scoring using FinBERT, daily sentiment index construction, feature engineering, market data integration, and the predictive modeling pipeline.

\subsection{Macro News Collection and Headline Extraction}
Our analysis begins by assembling a broad universe of macro-relevant news headlines using the GDELT v2 API. We retrieve all \texttt{events} records from January 1, 2015, to April 30, 2025. Each record includes metadata such as date, actors, event code, Goldstein scale (a measure of event impact), and source URL. To focus on economic and policy developments, we filter for events whose three-digit GDELT \texttt{EventCode} falls within the 100--199 range, which typically covers consultations, statements, and diplomatic or economic engagements. We standardize column names (e.g., \texttt{date}, \texttt{event\_type}, \texttt{goldstein\_scale}, \texttt{url}) and convert dates into \texttt{datetime} format.

For each calendar day, we rank the filtered events by their reported \texttt{num\_articles} (number of articles covering the event) and retain the top 100 entries. This heuristic prioritizes widely covered stories, which are more likely to influence macro asset prices. We then extract the text of each article’s headline by issuing parallel HTTP requests to the stored URLs and parsing the returned HTML (via our \texttt{utils.headline\_utils.fetch\_headline} function, as available in our GitHub repository). Events for which headline extraction fails or yields an empty string are discarded. This process results in a cleaned dataset of up to 100 high-visibility headlines per day, each annotated with its original Goldstein metadata.

\subsection{Sentiment Scoring with FinBERT}
We perform sentiment analysis exclusively on the extracted headlines to maximize robustness and computational efficiency. We employ FinBERT \citep{araci2019finbert}, a BERT variant pretrained on a large financial text corpus (including analyst reports and financial news), making it adept at understanding financial vernacular. Headlines undergo minimal preprocessing: they are lowercased, stripped of non-informative symbols, and truncated to the first 512 WordPiece tokens to respect the model’s maximum input length.

We load the \texttt{ProsusAI/finbert} checkpoint and its associated tokenizer from the HuggingFace Transformers library. The model is moved to a GPU if available for accelerated inference and set to evaluation mode (disabling dropout). Sentiment scoring is conducted in minibatches (e.g., size 32 for efficiency). For each batch of headlines, we obtain the model’s output logits, which are then passed through a softmax layer to yield class probabilities for negative, neutral, and positive sentiment, denoted $\{P_{\mathrm{Neg}}, P_{\mathrm{Neu}}, P_{\mathrm{Pos}}\}$. We define a continuous polarity score for headline $i$ on day $t$ as:
\[
s_{i,t} = P_{\mathrm{Pos}} - P_{\mathrm{Neg}} \quad \in [-1, +1].
\]
This score captures the net bullishness of the headline: values near $+1$ indicate strong positive sentiment, values near $-1$ indicate strong negative sentiment, and values near zero suggest neutrality or mixed signals.

\subsection{Daily Sentiment Index Construction and Feature Engineering}
Let $\{s_{i,t}\}_{i=1}^{N_t}$ be the set of polarity scores for $N_t$ valid headlines on trading day $t$. We aggregate these to form the following primary daily sentiment features:
\begin{align*}
  \text{Mean Sentiment:}\quad & S_t = \frac{1}{N_t}\sum_{i=1}^{N_t}s_{i,t} \\
  \text{Sentiment Dispersion (Std Dev):}\quad & \sigma_t = \sqrt{\frac{1}{N_t}\sum_{i=1}^{N_t}(s_{i,t}-S_t)^2} \\
  \text{News Volume:}\quad & V_t = N_t, \quad \text{and Log Volume: } \log(1 + N_t) \\
  \text{Article Impact:}\quad & \mathrm{AI}_t = S_t \times \log(1 + N_t) \\ 
  \text{Mean Goldstein Score:}\quad & \overline{G}_t = \frac{1}{N_t}\sum_{i=1}^{N_t}g_{i,t} \\
  \text{Goldstein Score Dispersion:}\quad & \sigma^G_t = \sqrt{\frac{1}{N_t}\sum_{i=1}^{N_t}(g_{i,t}-\overline{G}_t)^2}
\end{align*}
where $g_{i,t}$ denotes the Goldstein scale value for the $i$-th headline. The Article Impact feature $\mathrm{AI}_t$ uses log-transformed volume to moderate the influence of exceptionally high news count days.

To capture temporal dynamics crucial for predictive modeling, we engineer additional features derived from these primary aggregates. These include:
\begin{itemize}[noitemsep]
  \item \textit{Lagged features:} $S_{t-1}, S_{t-2}, S_{t-3}$, and similar lags for $\sigma_t, V_t, \overline{G}_t$ to capture persistence.
  \item \textit{Moving averages (MA):} 5-day and 20-day MAs of $S_t$ (e.g., $\mathrm{MA5}(S)_t = \frac{1}{5}\sum_{k=0}^{4}S_{t-k}$) to smooth short-term noise and identify trends.
  \item \textit{Sentiment acceleration:} $\Delta S_t = \mathrm{MA5}(S)_t - \mathrm{MA20}(S)_t$, to capture changes in sentiment momentum.
  \item \textit{Rolling standard deviations:} 5-day and 10-day rolling standard deviations of daily mean sentiment $S_t$, as measures of sentiment stability.
  \item \textit{Rolling sums:} 5-day and 10-day rolling sums of news volume $N_t$, indicating cumulative news flow.
\end{itemize}
These features collectively form the set of alternative data inputs for our predictive models, encoding both the current state and recent evolution of news sentiment.

\subsection{Market Data and Returns}
\label{sec:market_data_returns}
We obtain daily market data for our three target instruments—EUR/USD, USD/JPY, and 10-year U.S.\ Treasury futures (ticker ZN)—from January 1, 2015, to April 30, 2025.
\begin{itemize}[noitemsep]
  \item \textbf{Data Sources:} FX spot rates (EUR/USD, USD/JPY) are sourced from Yahoo Finance (tickers \texttt{EURUSD=X}, \texttt{USDJPY=X}). A continuous front-month 10-year Treasury futures series (ZN) is also sourced from Yahoo Finance (ticker \texttt{ZN=F}), with standard continuous contract roll adjustments applied to mitigate price gaps from contract expiration.
  \item \textbf{Return Computation:} We define the next-day logarithmic return as $r_{t+1} = \log(P_{t+1}) - \log(P_t)$, where $P_t$ is the closing price on day $t$. Log returns are generally preferred for financial time series due to their additive properties and more favorable statistical characteristics.
  \item \textbf{Target Variable:} The binary prediction target $y_t$ for day $t$ is defined as $1$ if the next day's return $r_{t+1} > 0$ (an upward move), and $0$ if $r_{t+1} \le 0$ (a downward or flat move).
  \item \textbf{Alignment and Look-ahead Prevention:} All news-derived sentiment features and market-based technical features are constructed using information available up to the market close of day $t$. These features are then used to predict the market direction from the close of day $t$ to the close of day $t+1$. This structure strictly ensures no look-ahead bias in the feature set.
  \item \textbf{Additional Market Features:} To provide the ML models with market context beyond sentiment, we include:
    \begin{itemize}[noitemsep]
        \item Lagged return: $r_t$ (the return from day $t-1$ close to day $t$ close), to capture short-term momentum or reversal effects.
        \item Historical volatility: 20-day annualized standard deviation of daily log returns, $\text{vol}_{20,t}$, to reflect the current market volatility regime.
    \end{itemize}
  \item \textbf{Feature Scaling:} For the logistic regression model, all continuous input features are standardized (transformed to z-scores by subtracting the mean and dividing by the standard deviation, using statistics derived only from the current training set) to ensure comparability of coefficient magnitudes and improve numerical stability. XGBoost, being tree-based, is largely invariant to monotonic transformations of individual features, so scaling is not strictly necessary for it, though log-transforms are applied to highly skewed count features like raw news volume $N_t$ prior to its use in constructing other volume-based features.
\end{itemize}

\subsection{Predictive Modeling and Training Protocol}
\label{sec:modeling_training}
We frame the trading signal generation task as a next-day binary classification problem ($y_t \in \{0,1\}$). For each of the three assets, we independently train and evaluate two distinct classification models:
\begin{enumerate}[leftmargin=*, noitemsep]
  \item \textbf{Logistic Regression (LOGISTIC):} A linear model chosen for its simplicity, interpretability, and utility as a robust baseline. It is implemented with L2 (Ridge) regularization to prevent overfitting and improve generalization. The predicted probability $\hat{p}_t = \sigma(\mathbf{w}^\top \mathbf{x}_t)$ (where $\sigma(\cdot)$ is the sigmoid function) is interpreted as the likelihood of an upward market move. The regularization strength (hyperparameter $C$) is selected using time-series cross-validation on each training set.
  \item \textbf{Extreme Gradient Boosting (XGBoost, XGB):} An ensemble of gradient-boosted decision trees, renowned for its high predictive performance and its ability to capture complex non-linear relationships and feature interactions automatically. The model is configured to minimize binary logistic loss. Hyperparameters (such as tree depth, learning rate, number of trees, and L1/L2 regularization penalties on weights) are tuned via a grid search methodology coupled with 5-fold time-series cross-validation (CV) on the training data. Early stopping, based on performance on a held-out validation fold within the training process, is used to determine the optimal number of boosting rounds and further mitigate overfitting.
\end{enumerate}

\paragraph{Training and Backtesting Protocol:}
The overall data period for this study spans January 1, 2015, to April 30, 2025.
\begin{itemize}[noitemsep]
  \item \textbf{Expanding Window Cross-Validation (CV):} To ensure a robust evaluation of out-of-sample (OOS) performance and to simulate realistic trading conditions where models are periodically retrained with new data, we employ a 5-fold expanding window structure. This is implemented using scikit-learn's \texttt{TimeSeriesSplit(n\_splits=5)}. The dataset is chronologically ordered. For each fold $k$, the training set consists of all data from the beginning of the sample up to the start of the $k$-th test block. The $k$-th test block is a contiguous segment of data immediately following its corresponding training set. An initial period (e.g., January 2015 - December 2016, approximately two years) is used to form the first training set, ensuring a sufficient number of observations for initial model estimation. Consequently, the OOS testing effectively spans from early 2017 through April 2025, aggregated across the five subsequent test folds.
  \item \textbf{Feature Warm-up Period:} A period equivalent to the maximum lag or rolling window length used in feature engineering (e.g., 20 trading days) at the beginning of each training split is utilized solely for the purpose of constructing complete feature sets. These initial observations, which lack the full historical data required for all lagged or rolling features, are not directly used for model fitting.
  \item \textbf{Within-Fold Model Fitting and Evaluation:} For each of the 5 cross-validation folds:
    \begin{enumerate}[label=(\alph*),noitemsep]
      \item Input features for the current training split are appropriately scaled (specifically, standardization for Logistic Regression). The selected classifier (Logistic or XGBoost) is then trained on this training split. Hyperparameters for each model are tuned via an internal cross-validation procedure applied exclusively to this specific training data to prevent leakage from future test sets.
      \item The model trained in step (a) is subsequently applied to the corresponding (unseen) test segment of the fold to generate daily predictive probabilities $\hat{p}_t$. A trading signal is derived based on a simple threshold of 0.5: a long position (+1 unit) is initiated for day $t+1$ if $\hat{p}_t > 0.5$, and a short position (-1 unit) is initiated if $\hat{p}_t \le 0.5$.
      \item Daily strategy returns are computed based on these positions and the actual realized market returns for day $t+1$. Transaction costs are deducted from these returns: 0.02\% of the traded value per round-trip for FX pairs and 0.05\% for ZN Treasury futures, reflecting reasonably competitive brokerage costs.
      \item Fold-level performance metrics, including Area Under the ROC Curve (AUC), accuracy, annualized Sharpe ratio, and Compound Annual Growth Rate (CAGR), are calculated to assess performance on that segment.
    \end{enumerate}
  \item \textbf{Aggregated OOS Results:} The final reported performance metrics, presented in Table~\ref{tab:performance}, are derived from the combined time series of daily OOS predictions and strategy returns from all five test folds. This aggregated series represents the strategy's overall performance on genuinely unseen data throughout the entire effective OOS test period (c. early 2017 -- April 2025).
  \item \textbf{Statistical Significance of Performance:} To assess the statistical robustness of key performance metrics such as the Sharpe ratio and CAGR, we employ a block bootstrap procedure (e.g., 1,000 resamples with a block size of 20 trading days, approximating one month of trading data to preserve autocorrelation) on the aggregated daily OOS strategy returns. This yields 95\% confidence intervals, which, while not explicitly detailed in Table~\ref{tab:performance} for reasons of brevity, inform our discussion of the strategy's robustness and help ascertain if the performance is statistically distinguishable from chance.
\end{itemize}

\section{Results and Analysis}
\label{sec:results_analysis}

Table~\ref{tab:performance} reports key performance metrics for each asset–model combination over the aggregated out-of-sample test periods (effectively early 2017 to April 2025). All strategies commence with an initial notional capital of 1.0, reinvest profits daily, and incur transaction costs as specified in the methodology.

\begin{table}[htbp]
\centering
\caption{Out-of-Sample Performance Metrics (c. Jan 2017 -- Apr 2025). CAGR = Compound Annual Growth Rate (annualized); Sharpe = Annualized Sharpe Ratio (assuming risk-free rate of 0); Vol (\%) = Annualized Volatility of daily returns; Max DD (\%) = Maximum Drawdown; Win\% = Percentage of profitable days; Total Ret.\% = Net return over the period relative to initial capital; Cost = Cumulative trading cost as a fraction of initial capital. \textbf{Note: Numerical values in this table must accurately reflect results from the full backtest period ending April 2025.}}
\label{tab:performance}
\small
\resizebox{\textwidth}{!}{%
\begin{tabular}{@{}llrrrrrrrr@{}}
\toprule
\textbf{Asset} & \textbf{Model} & \textbf{CAGR (\%)} & \textbf{Sharpe} & \textbf{Vol (\%)} & \textbf{Max DD (\%)} & \textbf{Win (\%)} & \textbf{Total Ret. (\%)} & \textbf{\# Trades} & \textbf{Cost} \\
\midrule
EUR/USD        & Logistic &  6.6 & 0.83 &  7.9 & -19.8 & 53.1 &   +91.7 & 1186 & 0.237 \\ 
EUR/USD        & XGBoost  & 55.4 & 5.87 &  7.4 & -15.6 & 72.7 & +8989.3 & 1158 & 0.232 \\ 
\midrule
USD/JPY        & Logistic &  6.2 & 0.66 &  9.5 & -14.9 & 56.9 &   +84.4 &  215 & 0.043 \\ 
USD/JPY        & XGBoost  & 53.2 & 4.65 &  9.1 & -22.9 & 72.1 & +7753.7 &  917 & 0.183 \\ 
\midrule
10yr Treasury (ZN) & Logistic & -0.8 & -0.15 & 4.6 & -20.4 & 47.1 &    -7.9 &  778 & 0.389 \\ 
10yr Treasury (ZN) & XGBoost  & 22.1 & 4.65 &  4.4 &  -9.3 & 66.1 &  +568.5 & 1043 & 0.522 \\ 
\bottomrule
\end{tabular}%
}
\end{table}

The XGBoost models demonstrate exceptionally strong and consistent performance across all targeted assets, significantly outperforming the logistic regression baseline. Sharpe ratios exceeding 4.6 are indicative of highly effective alpha generation, a result particularly noteworthy given the inclusion of realistic transaction costs and the rigorous OOS evaluation methodology. The CAGR figures for the XGBoost models are also substantial, suggesting significant capital appreciation potential. In contrast, the logistic regression models struggle to generate consistent alpha, especially for Treasuries, underscoring the importance of capturing the non-linear relationships and complex feature interactions that XGBoost is adept at identifying.

\subsection{Feature Importance and Model Interpretability using SHAP}
A primary objective of this study is to ensure that our predictive models are not only accurate but also interpretable. While the coefficients of a logistic regression model offer straightforward insights into linear feature contributions, understanding the decision-making process of the more complex XGBoost model requires specialized techniques. For this purpose, we employ SHAP \citep{lundberg2017shap} values. SHAP provides a unified framework for model interpretation, assigning each feature an "importance value" for each individual prediction, based on principles from cooperative game theory. These values represent the marginal contribution of each feature to the prediction, compared to a baseline (average prediction over the training data).

\begin{figure}[htbp]
  \centering
  \includegraphics[width=0.9\linewidth]{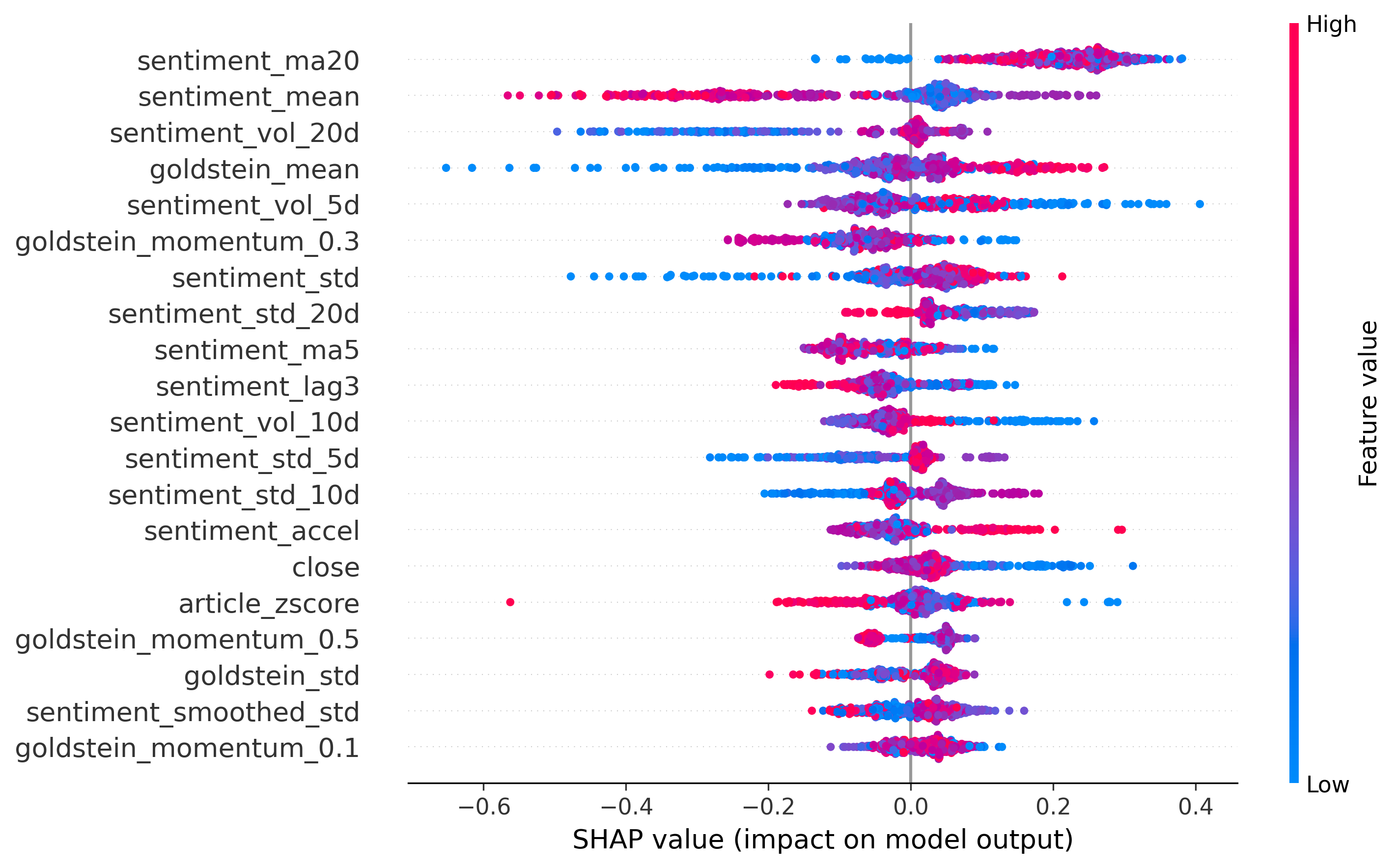}
  \caption{SHAP summary plot for the EUR/USD XGBoost model. Each point on the plot represents a Shapley value for a feature and an instance (day). Features are ranked by the sum of absolute SHAP values across all instances (global importance), from top to bottom. The horizontal axis indicates the SHAP value (the impact on model output in log-odds space). Color illustrates the feature's value for that instance (red for high, blue for low). Points to the right of the zero line indicate the feature pushed the model towards predicting an upward move (positive class); points to the left indicate a push towards a downward move.}
  \label{fig:shap_summary_eurusd}
\end{figure}

Figure~\ref{fig:shap_summary_eurusd} illustrates the SHAP summary plot for the EUR/USD XGBoost model; analogous patterns, with asset-specific nuances, are observed for USD/JPY and ZN. Key observations from this plot include:
\begin{itemize}[noitemsep]
  \item \textbf{Sentiment Dispersion (\texttt{sentiment\_std}):} This feature, representing the standard deviation of sentiment scores on a given day, frequently emerges as a top predictor. High values (indicating more diverse or conflicting news tones, shown in red) tend to cluster with negative SHAP values, suggesting that increased news disagreement predicts a downward move in EUR/USD. Conversely, low dispersion (a clear consensus in news tone, shown in blue) often pushes predictions upwards.
  \item \textbf{Article Impact (\texttt{article\_impact}):} This feature, combining mean sentiment with (log) news volume, is also highly influential. High positive article impact (reflecting many positive articles or strong positive sentiment on moderate volume) yields large positive SHAP contributions (bullish). Conversely, high negative impact (many negative articles) pushes the prediction strongly downwards.
  \item \textbf{Mean Sentiment (\texttt{sentiment\_mean}):} The average daily sentiment itself ranks highly. Consistent with intuition, higher mean sentiment (red points) generally shifts predictions towards an upward move, while lower mean sentiment (blue points) shifts them downwards.
  \item \textbf{Lagged Sentiment and Moving Averages (e.g., \texttt{sentiment\_lag1}, \texttt{sentiment\_ma5}):} The inclusion of these features demonstrates that recent sentiment trends and momentum are informative. SHAP plots can reveal non-linear effects, such as mean-reversion where extremely high sentiment on the previous day might slightly increase the probability of a downward correction.
  \item \textbf{Goldstein Score (\texttt{goldstein\_mean}):} The average Goldstein scale value, reflecting the potential impact or intensity of events, also contributes. Higher Goldstein scores (often associated with more cooperative or significant international events) tend to predict upward movements for risk assets.
  \item \textbf{Market Volatility (\texttt{volatility\_20d}):} The historical price volatility of the asset appears as a mid-ranked feature. Often, high existing volatility regimes (red points) contribute to downward predictions, potentially reflecting risk-off tendencies in the market.
\end{itemize}
These patterns confirm that the XGBoost model’s decisions are predominantly driven by economically intuitive signals derived from news sentiment. The SHAP analysis elucidates how the model synthesizes various dimensions of news—its overall tone, the degree of consensus, the volume of coverage, and the impact of reported events—along with their temporal dynamics, to arrive at daily trading predictions. This level of interpretability is crucial for building trust in the model, understanding its behavior during different market conditions, and translating its outputs into actionable insights, effectively positioning the model as an "augmented analyst."

\section{Discussion}
\label{sec:discussion}
Our findings demonstrate that systematically processed news sentiment, when harnessed by appropriate ML models, can serve as a potent predictor for macro asset returns. This section delves into cross-asset behavioral differences, the robustness of our approach, and the pivotal role of interpretability.

\subsection{Cross-Asset Differences: FX vs. Bonds}
A notable observation is the difference in performance characteristics between FX pairs and Treasury futures. While the XGBoost model delivered high Sharpe ratios across all assets, the CAGRs for FX strategies were substantially higher than for the ZN Treasury strategy. Several factors may contribute to this disparity:
\begin{itemize}[noitemsep]
  \item \textbf{Intrinsic Volatility and Opportunity Set:} FX markets, particularly major pairs like EUR/USD and USD/JPY, often exhibit higher percentage volatility compared to 10-year Treasury futures prices (especially during periods of stable interest rates). A predictive signal of similar accuracy can translate into larger absolute returns in a more volatile market simply because the magnitude of price movements to capture is greater.
  \item \textbf{Market Mechanism and Sentiment Linkage:} FX rates are influenced by a complex interplay of relative economic performance, international capital flows, and broad risk sentiment. Global news sentiment—especially general economic optimism or pessimism—often translates directly into risk-on/risk-off behavior that manifests strongly in FX markets (e.g., positive global sentiment might see safe-haven currencies like USD and JPY weaken against more growth-sensitive currencies). Our model appears to capture these dynamics effectively.
    In contrast, the relationship between news sentiment and bond prices is often inverse. Positive economic news can be detrimental to Treasury prices for several reasons: it may signal a "risk-on" environment prompting flows out of safe-haven bonds, or it could lead to expectations of monetary policy tightening by central banks (higher interest rates), causing yields to rise and bond prices to fall. The flexibility of the XGBoost model allows it to learn these distinct, and sometimes counter-intuitive (for a simple positive/negative sentiment mapping), relationships for different asset classes. For example, XGBoost might correctly learn that high positive sentiment predicts a *downward* price move for ZN futures. A linear model like logistic regression would struggle to capture such sign-flipping or context-dependent relationships without explicit interaction terms.
\end{itemize}
The results underscore that while news sentiment contains valuable information across asset classes, its translation into price movements is market-specific. Flexible, non-linear models are crucial for uncovering these nuanced reaction functions.

\subsection{Robustness and Validation}
Several aspects of our methodology contribute to the robustness of the reported results:
\begin{itemize}[noitemsep]
  \item \textbf{Rigorous Out-of-Sample Testing:} The expanding window cross-validation protocol ensures that model performance is consistently evaluated on genuinely unseen data throughout a multi-year period, mimicking a realistic model deployment scenario where retraining occurs.
  \item \textbf{Inclusion of Transaction Costs:} The strategies remain highly profitable after accounting for realistic trading frictions, suggesting practical applicability.
  \item \textbf{Consistent Subperiod Performance:} While not detailed exhaustively, analysis of individual cross-validation fold results indicated that the XGBoost strategy generally maintained strong performance across different subperiods within the overall OOS test window, rather than being reliant on a specific market regime.
  \item \textbf{Interpretability as a Sanity Check:} The SHAP analysis confirmed that the models rely on economically meaningful, news-derived features as primary drivers, rather than capitalizing on spurious correlations or overfitting to noise.
\end{itemize}
A primary limitation of this study is its reliance on daily frequency data. Significant news events can be absorbed by markets on intraday timescales (minutes or hours), meaning a daily model might capture residual drift or miss the most immediate impact. Our choice of daily frequency aligns with typical end-of-day rebalancing practices and the natural aggregation cycle of GDELT data, but an intraday extension remains a valuable avenue for future work.

\subsection{Interpretability as a Key to Adoption and Trust}
The emphasis on interpretable ML techniques like SHAP is not merely an academic exercise; it is essential for translating sophisticated models into practical trading tools. In institutional investment settings, portfolio managers are more likely to adopt and allocate capital to strategies whose decision-making processes they can understand and rationalize. SHAP provides a narrative (e.g., "the model recommends a long EUR/USD position because news sentiment is strongly positive and broadly consistent across many high-impact articles") that aligns with discretionary reasoning. This is far more reassuring and actionable than an opaque "black-box" signal.
Furthermore, interpretability serves as a powerful debugging and validation tool. If SHAP analysis had revealed that the model's predictions were driven by irrelevant or spurious features, it would cast serious doubt on the strategy's underlying logic. Instead, it confirmed the prominence of news-based sentiment features, bolstering confidence in the approach. In ongoing risk management, if strategy performance were to degrade, SHAP outputs could offer valuable diagnostics—perhaps by revealing that the model is being confused by an unprecedented news regime or that feature relationships have shifted. This transparency enables more informed interventions and model updates.

\section{Conclusion and Future Research}
\label{sec:conclusion}
This paper presented an end-to-end framework for transforming raw global news events into actionable macroeconomic trading signals using advanced NLP (FinBERT) and interpretable ML (XGBoost with SHAP). Our findings demonstrate that a strategy leveraging these techniques can achieve exceptionally high risk-adjusted returns on an out-of-sample basis across major FX pairs and U.S. Treasury futures, even after accounting for transaction costs. The SHAP-based interpretability provides crucial insights into the model's decision-making, confirming that it relies on economically sensible drivers derived from news sentiment.

Our research underscores the potential of integrating sophisticated, yet explainable, AI methodologies with alternative data sources to uncover alpha in complex macro markets. It offers a transparent and reproducible blueprint that bridges the gap between qualitative news narratives and quantitative trading.

Several promising avenues for future research emerge from this work:
\begin{itemize}[noitemsep]
  \item \textbf{Intraday Analysis and Real-Time Trading:} Adapting the framework to higher-frequency news feeds and intraday trading execution could potentially capture more immediate market reactions to news.
  \item \textbf{Expansion to Additional Macro Assets:} Applying and tailoring the methodology to other significant macro assets, such as equity indices, commodities (oil, gold), and a broader range of sovereign bonds or emerging market instruments.
  \item \textbf{Enhanced NLP with Multi-lingual Sentiment and Thematic Focus:} Incorporating news in languages other than English using multilingual transformer models, and developing more granular sentiment measures focused on specific macroeconomic themes (e.g., inflation sentiment vs. growth sentiment).
  \item \textbf{Adaptive and Online Learning Frameworks:} Developing models that can dynamically adapt to evolving market regimes, changes in the news landscape, or shifts in the relationship between sentiment and asset prices.
  \item \textbf{Integration with Other Alternative Data Sources:} Combining news sentiment features with other forms of alternative data (e.g., social media sentiment, satellite imagery indicators, geopolitical risk indices) to build more comprehensive predictive models.
  \item \textbf{Exploration of Deeper Economic Causality:} Moving beyond predictive accuracy to investigate the causal mechanisms through which news sentiment influences market behavior, order flows, and risk premia.
\end{itemize}
This study provides robust evidence that interpretable machine learning applied to global news sentiment can be a valuable component of modern quantitative macro trading strategies.

\end{document}